\documentclass[conference]{IEEEtran}

\usepackage{booktabs} % For formal tables
\usepackage{url}
\usepackage[utf8x]{inputenc}
\usepackage{amsmath}
\usepackage{unicode}
\usepackage{tikz}

\newcommand{\REPO}{\url{https://github.com/endgameinc/homoglylph}}

\ifCLASSINFOpdf
\else
\fi
\begin{document}
\title{Detecting Homoglyph Attacks with a Siamese Neural Network}

\author{\IEEEauthorblockN{Jonathan Woodbridge, Hyrum S. Anderson, Anjum Ahuja, Daniel Grant}
\IEEEauthorblockA{Endgame\\
Email: \{jwoodbridge, hyrum, aahuja, dgrant\}@endgame.com}}

\maketitle

\begin{abstract}
A homoglyph (name spoofing) attack is a common technique used by adversaries to obfuscate file and domain names.  This technique creates process or domain names that are visually similar to legitimate and recognized names.  For instance, an attacker may create malware with the name \textit{svch0st.exe} so that in a visual inspection of running processes or a directory listing, the process or file name might be mistaken as the Windows system process \textit{svchost.exe}.  There has been limited published research on detecting homoglyph attacks. Current approaches rely on string comparison algorithms (such as Levenshtein distance) that result in computationally heavy solutions with a high number of false positives. In addition, there is a deficiency in the number of publicly available datasets for reproducible research, with most datasets focused on phishing attacks, in which homoglyphs are not always used.%Some authors have attempted to improve false positive rates by manually crafting distance functions that take visual similarity into account.  However, such techniques require the creation of an all-to-all similarity score between characters.  This is an intractable solution especially when considering that process names may be composed of unicode characters.

This paper presents a fundamentally different solution to this problem using a Siamese convolutional neural network (CNN).  Rather than leveraging similarity based on character swaps and deletions, this technique uses a learned metric on strings rendered as images: a CNN learns features that are optimized to detect visual similarity of the rendered strings. The trained model is used to convert thousands of potentially targeted process or domain names  to feature vectors.  These feature vectors are indexed using randomized KD-Trees to make similarity searches extremely fast with minimal computational processing.  This technique shows a considerable 13\% to 45\% improvement over baseline techniques in terms of area under the receiver operating characteristic curve (ROC AUC).  In addition, we provide both code and data to further future research.

%.  Feature vectors are indexed using KD-Trees and each time a new file is created or a domain name is observed, its name is converted to a feature vector and searched in the index for other similar names.  Anything with a Euclidean distance below a threshold is flagged as a potential spoofing attack. Our method is validated using two datasets, one with process names and one with domain names, and is compared to state-of-the-art techniques.  

%More specifically, we derive a training set of pairs; process names that are name spoofing and those that are not.  Each pair of process names is converted to a pair of images and passed to a Siamese convolutional neural network.  The network is trained to convert each image to a feature vector such that the euclidean distance between a spoofing attack and its targeted name is $0.0$ while the distance between two process names that are not a spoofing attack is $1.0$.  We use 

%  This system is currently in our product deployed to thousands of machines.
\end{abstract}


\section{Introduction}
\label{introduction}
Cyber attackers have long leveraged creative attacks to infiltrate networks.  One simple attack uses homoglyphs or name spoofing to obfuscate malicious purpose.  These attacks occur for both domain names and process names.  Attackers may use simple replacements such as \textit{0} for \textit{o}, \textit{rn} for \textit{m}, and \textit{cl} for \textit{d}. Swaps may also include Unicode characters that look very similar to common ASCII characters such as \textit{ł} for \textit{l}. Other attacks append characters to the end of a name that seem valid to a user such as \textit{svchost32.exe}, \textit{svchost64.exe}, and \textit{svchost1.exe}.  The hope is that these processes or domain names will go undetected by users and security organizations by blending in as legitimate names.  %For process names, attackers can drop visually similar binaries in the same directory as their counterparts making them appear almost identical to an observer. 

One naive approach for discovering name spoof attacks is to calculate the edit (Levenshtein) distance of each new process or domain name to each member of a set of processes or domain names to monitor.  In general, edit distance measures the number of \textit{edits} (insertions, deletions, substitutions or transpositions) to convert one string to another. A distance less than or equal to a pre-defined threshold is flagged as a potential spoof.  In practice, this approach suffers from a poor False Positive (FP)/False Negative (FN) tradeoff. %In addition, if attackers discover the threshold, they can craft spoofing attacks to always be greater than the threshold.  For example, if the threshold is set to an edit distance of 2, then an attacker will make sure that all spoofing names are at least edit distance 3 from the process name they are spoofing.  

Another approach is to create a custom edit distance function that accounts for the visual similarity of substitutions, so that substituting a character with a visually similar character result in a smaller edit distance than a visually distinct character \cite{linari2009typo, EditDistanceVC}.  As shown later in the paper, these techniques result in only modest improvements over standard edit distance functions.  In addition, these techniques are in large part manually crafted, making them very difficult to enumerate and maintain, especially when considering the full Unicode alphabet.

To overcome the shortcomings of the aforementioned methods, this paper presents a metric-learning technique based on a Siamese convolutional neural network (CNN). A training set $\left\{ \left(s_i, s^\prime_i, y_i\right)\right\}_{i=1}^n$ is composed of $n$ pairs of strings consisting of either process names or domain names, together with a distance target (similarity label) $y_i$.  A pair of strings $(s_i, s^\prime_i)$ for which $s^\prime_i$ is a spoof of $s_i$ (or vice versa), we assign $y_i=0$ (similar), and $y_i=1$ (dissimilar) otherwise.  Each string $s_i$ and its pair $s^\prime_i$ is then rendered as a binary image $\mathbf{x}_i$ and $\mathbf{x}^\prime_i$, respectively.  The Siamese CNN is explicitly trained to convert images to features vectors such that the distance between feature vectors of spoofing pairs target a distance of $0.0$, and at least $1.0$ otherwise. The model is deployed as a defensive measure as follows.  We convert all common or potentially targeted domain or process names to feature vectors.  These feature vectors are indexed using a randomized KD-Tree index.  When a new process or domain name is observed, it is converted to a feature vector by the CNN and searched in the KD-Tree index to find any visually similar matches.  If a match exists, then a homoglyph attack is detected.  

%This system is currently in production and runs on process logs of tens of thousands of hosts.  The system is extremely scalable with a low false positive rate.  Our current system has a $1/10000$ false positive rate.  In addition, we've implemented a whitelisting features that allows us to avoid previously seen FPs making our effective false positive rate much lower.  

On the surface, this problem may seem similar to other well studied problems.  For example, there is a large body of work that addresses the discovery of phishing attacks \cite{ma2009identifying}.  Often these attacks are waged via email so as to trick unsuspecting victims to click on malicious domain names that appear to be benign in order to steal information.  Despite some similarities, much of the work in phishing detection is largely not applicable to detecting homoglyph attacks.  First, phishing attacks often use domains that appear to be legitimate, but are visually distinct from the benign domain that they are impersonating. For example, an attacker may register the domain \textit{google-weekly-updates.com}.  In this example, the domain is very different from \textit{google.com} and probably unlikely to be registered (at least not at the time of this publication!).  In fact, previous works found that the likelihood of a phishing attack grows with increasing edit distance between the phishing domain and the legitimate domain \cite{ma2009identifying}.  Second, phishing detection can use contextual information such as appearance of the web pages (e.g., does the content of \textit{fake-facebook.com} appear legitimate?) and whois information (i.e., registration information related to the domain name). % Unfortunately, reliable contextual information such as this is not readily available for all applications (such as process names).  For instance, anti-virus detection (AV) can provide contextual information, however, AV does not catch all threats and is regularly beat with new more advanced malware.  Furthermore, malware trends exhibit no signs of decreasing \cite{securityreport}, and name spoofing is a viable stealth mechanism until reactive signature-based anti-virus solutions are updated.

There has also been a large amount of work in regards to finding nearest string matches in other domains such as data cleansing, spell checking, and bioinformatics \cite{deng2013top, wang2011fast, deng2014pivotal}. However, those works did not consider visual similarity of characters and do not apply to the problem at hand.  Instead, this work is largely inspired by the work in \cite{hadsell2006dimensionality} that uses a similar Siamese network used in this paper to classify digits in the MNIST \cite{lecun1998gradient} and NORB \cite{lecun2004learning} datasets.
 
Our work makes three primary contributions:

\begin{enumerate}
  \item Presents a generic name spoofing detection technique using a Siamese CNN, which to our knowledge, is the first such application of metric learning to homoglyph detection,
  \item Compares the system's efficacy to other common string comparison techniques, and 
  \item Contributes source code to reproduce results in this paper as well as two datasets to further research in this area\footnote{\REPO}.
\end{enumerate}

In Section \ref{related_work} we take a deeper dive into related work as well as the motivations behind this work.  In Section \ref{method} we discuss the high-level design of the system and the architecture of our neural network.  In Section \ref{results} we compare our neural network to other string matching techniques.  We conclude the paper in Section \ref{conclusion} with some closing thoughts. 

\section{Related Work}
\label{related_work}
An extensive amount of work has been devoted to efficient string matching.  Some of this work is focused on making string matching fast \cite{deng2014pivotal, deng2013top} while other work focuses on improving the quality of nearest neighbor searches \cite{wang2011fast}.  However, conventional string matching algorithms are not an effective technique for detecting name spoofing.  For example, consider the windows application \textit{iexplore.exe}.  A malicious user may create a piece of malware with the name \textit{iexp1orc.exe} that is an edit distance of $2$ from the original executable.  In this case, a system that labels all process names with an edit distance of $2$ or less would catch this spoof attack.  However, consider also the common windows process \textit{explorer.exe}.  This process is also an edit distance $2$ from \textit{iexplore.exe} resulting in a false positive.  

The key to detecting name spoofing attacks is to make visual comparisons. When visually comparing the three strings \textit{iexplore.exe}, \textit{iexp1orc.exe}, and \textit{explorer.exe}, one of the strings looks very different from the other two.  The first and last character (before \textit{.exe}) of the string \textit{explorer.exe} has a different shape from the other two strings making it very distinguishable.  Such distinguishable characters are unlikely to fool anyone in a spoofing attack, but are lost in basic string matching systems.  

There are many subtle string updates that result in a string that appears almost identical to the original string.  In addition, the Windows operating system supports Unicode characters resulting in an exponentially large number of string swaps making signature-based detection infeasible (i.e., a lookup table of all possible character swaps).  Several spoofing attempts are given in Table \ref{table:spoof_examples1}.  Notice how easily spoofing strings may be overlooked. Authors in \cite{trabasso2014orthography} give more in-depth analysis of characteristics that make strings appear visually similar.

\begin{table}[h!]
\centering
  \caption{Example of process name homoglyphs}
  \label{table:spoof_examples1}  
  \begin{tabular}{ l | l | c }
   	Original & Spoof & Edit Distance \\ 
   	\hline
    \hline
    \textit{SVCHOST.EXE} & \textit{SVCH0ST.EXE} & 1 \\
    \textit{LSASS.EXE} & \textit{LS4SS.EXE} & 1 \\
    \textit{iexplore.exe} & \textit{iexp1orc.exe} & 2 \\
    \textit{chtime.exe} & \textit{chtirne.exe} & 2 \\
  \end{tabular}
\end{table}

Authors in \cite{EditDistanceVC,linari2009typo} attempted to improve upon conventional edit distance functions by adding knowledge of visual likeness in characters.  For example, swapping a \textit{r} for a \textit{n} would result in a smaller distance than a \textit{y} or a \textit{b}.  This technique relied on a largely manual step of deriving similarity measures between characters and did not include the massive unicode set.  While this method generally improves upon conventional techniques, it still exhibits a high false positive rate.

\subsection{Phishing}
Phishing attacks can be broken down into four categories \cite{garera2007framework}: 

\begin{enumerate}
  \item Obfuscating a domain name with an IP address,
  \item Obfuscating a domain name with another domain name, 
  \item Obfuscating a domain name within a longer domain name, and
  \item Obfuscating a domain name using misspellings and common typos.
\end{enumerate}

Examples of each type of phishing attack is given in Table \ref{table:phish_examples2}. The first three obfuscation techniques result in domains that are not visually similar.  While the target domain name may be a substring of the phishing domain name, the two strings are visually different.  The fourth obfuscation technique may seem to be similar to process name spoofing, however, misspellings and typos are not necessarily visually similar.  

\begin{table}[h!]
\centering
  \caption{Example of URL phishing attacks. Only the last is a homoglyph.}
  \label{table:phish_examples2}
  \begin{tabular}{ c | l | l }
   	Type & Phish URL & Target Domain \\ 
   	\hline
    \hline
    1 & \textit{202.0.0.1/google.com} & \textit{google.com} \\
    2 & \textit{badDomain.com/google.com} & \textit{google.com} \\
    3 & \textit{google.com.badDomain.com} & \textit{google.com} \\
    4 & \textit{google.om} & \textit{google.com} \\
  \end{tabular}
\end{table}

Machine learning based approaches for detecting phishing domains rely on two types of features \cite{garera2007framework, zhang2007cantina, basnet2008detection, xiang2011cantina+, marchal2016know}.  These include \textit{domain-based features} that are derived directly from the domain name and \textit{page-based features} that are derived from the hosted page.

These techniques have been effective in phishing detection, however, they do not focus on visual similarity.  In fact, authors in \cite{ma2009identifying} found that the likelihood of a phishing attack grows with increasing edit distance between the phishing domain and the legitimate domain.  Thus, methods to detect phishing attacks are largely not applicable to detecting spoofing attacks.  While a new set of features could be derived to detect name spoofing, this process is extremely time consuming and highly susceptible to the cat and mouse games waged by adversarial actors.  For these reasons, this work focused solely on visual appearance and relies on convolutional neural networks to derive its own visual features.

%Most of these features are used in phishing attacks are based on contextual information that are not available in all name spoofing domains (such as process names).  While a new set of features could be derived to detect name spoofing, this process is extremely time consuming and highly susceptible to the cat and mouse games waged by adversarial actors.  For these reasons, this work focused solely on visual appearance and relies on the power of convolutional neural networks to derive its own visual features.

%LЅΑЅЅ.ΕΧΕ
%LЅΑЅЅ.ΕΧΕ
\subsection{Siamese Neural Networks}
Siamese neural networks were first introduced in 1993 by Bromely and LeCun as a method to validate handwritten signatures \cite{bromley1993signature}.  At its core, a Siamese neural network is simply a pair of identical neural networks (i.e., shared weights) which accept distinct inputs, but whose outputs are merged by a simple comparative energy function.  The key purpose of the neural network is to map a high-dimensional input (e.g., an image) into a target space, such that a simple comparison of the targets by the energy function approximates a more difficult-to-define ``semantic'' comparison in the input space.

Mathematically, if a neural network $\mathbf{g}_{\mathbf{W}} : \mathbb{R}^n \mapsto \mathbb{R}^d$ is parameterized by weights $\mathbf{W}$, and we choose simple Euclidean distance for our comparative energy function 
$E: \mathbb{R}^d \times  \mathbb{R}^d \mapsto \mathbb{R}$, then the Siamese network computes dissimilarity between the pair of images $\left(\mathbf{x}_1, \mathbf{x}_2\right)$ simply as
\begin{align}
d_{\mathbf{W}}\left(\mathbf{x}_1, \mathbf{x}_2 \right) &= E\left( \mathbf{g}_{\mathbf{W}}\left(\mathbf{x}_1\right), \, \mathbf{g}_{\mathbf{W}}\left(\mathbf{x}_2\right) \right) \nonumber\\
&= \, || \mathbf{g}_{\mathbf{W}}\left(\mathbf{x}_1\right) - \mathbf{g}_{\mathbf{W}}\left(\mathbf{x}_2\right) ||_2. \label{eqn:siamese}
\end{align}

Note that $\mathbf{g}_{\mathbf{W}}$ represents a family of functions parameterized by $\mathbf{W}$. We wish to learn $\mathbf{W}$ such that $d_{\mathbf{W}}\left(\mathbf{x}_1, \mathbf{x}_2 \right)$ is small if $\mathbf{x}_1$ and $\mathbf{x}_2$ are similar, and large if they are dissimilar.  At first glance, one may be tempted to choose $\mathbf{W}$ simply minimizing $d_{\mathbf{W}}$ over pairs of similar inputs; however, this may lead to degenerate solutions such as $\mathbf{g}_{\mathbf{W}} = \mathrm{constant}$, for which $d_{\mathbf{W}}$ is identically zero.  Instead, previous research has employed \textit{contrastive loss} to ensure that similar inputs result in small $d_{\mathbf{W}}$, while simultaneously pushing $d_{\mathbf{W}}$ to be large for dissimilar inputs
\cite{chopra2005learning}.

Chopra et al. \cite{hadsell2006dimensionality} proposed  a contrastive loss function of the form
\begin{align}
\mathcal{L}\left(\mathbf{W}\right) &= \sum_{i=1}^P (1-y_i)L_S\left( d_{\mathbf{W}}^i  \right) + y_i L_D\left( d_{\mathbf{W}}^i \right),
\label{eqn:contrastive}
\end{align}
where $y_i=0$ if the images in the $i$th input pair $\left(\mathbf{x}_1, \mathbf{x}_2 \right)^i$ are deemed similar and $y_i=1$ if dissimilar, $d_{\mathbf{W}}^i = d_{\mathbf{W}}\left( \left(\mathbf{x}_1, \mathbf{x}_2 \right)^i \right)$ is the Siamese network dissimilarity for the $i$th pair, and the summation occurs over all $P$ input pairs.  The authors chose partial loss for similar pairs to be squared loss, $L_S(x) = x^2$, while partial loss for dissimilar pairs was chosen to be the squared hinge loss with margin $\alpha$, $L_D(x) = \left(\mathrm{max}\left\{0, \alpha - x\right\}\right)^2$.  Intuitively, this loss aims to shrink the distance between feature vectors of similar pairs to $0$, while expanding the distance between dissimilar pairs to be at least $\alpha$.  In our experiments, we use a margin of $\alpha=1$.

Since the loss function is differentiable with respect to $\mathbf{W}$, the weights can learned via backpropagation.  Notable is the fact that after the weights $\mathbf{W}$ have been trained, the network $\mathbf{g}_{\mathbf{W}}$ may be used in isolation to map from the space of images to the compact target feature space for simple comparison.

\subsection{Indexing Strings}

%\begin{figure}
%\center
%\begin{tikzpicture}[sibling distance=10em,
%  every node/.style = {shape=rectangle, rounded corners,
%    draw, align=center},
%    level 2/.style = {sibling distance=5em}]
%  \node {{[\textbf{0},5,3,4]}\\{[\textbf{2},5,2,1]}\\{[\textbf{6},5,4,5]}\\{[\textbf{3},5,2,2]}}
%    child { node {{[0,5,3,\textbf{4}]}\\{[2,5,2,\textbf{1}]}} 
%      child { node {{[2,5,2,1]}} }
%      child { node {{[0,5,3,4]}} }}
%    child { node {{[\textbf{6},5,4,5]}\\{[\textbf{3},5,2,2]}} 
%      child { node {{[3,5,2,2]}} }
%      child { node {{[6,5,4,5]}} } };
%\end{tikzpicture}
%\caption{Basic KD-Tree index built from four nodes.  The bold numbers highlight the split dimension.}
%\label{fig:kd_tree}
%\end{figure}

Once a Siamese neural network is trained to convert strings to a feature vector, we must select many process names (or domain names) that we are interested in monitoring (i.e., which names do we expect to be targeted in a spoof attack?).  This list is tractable as it is less likely for an attacker to spoof a process (or domain) name that is known by very few people.  However, this list can easily grow into the hundreds of thousands.  For example, someone interested in monitoring domain names may want to monitor the top 250K common domains around the world.  A naive approach is to compute the Euclidean distance between a suspect string's feature vector and every string's feature vector that is being monitored.  This brute force nearest search can be improved significantly using indexing.

We employ (randomized) KD-Trees as a geometrical index \cite{bentley1975multidimensional} to quickly search for similar feature vectors. There are several algorithms for performing nearest neighbor search \cite{jegou2011product, ge2013optimized, wang2014hashing, muja2014scalable}, and many may work for this technique.  KD-Trees were chosen for their simplicity and availability of open source tools.  

% Hyrum summarizes some points in the following paragraphs
In KD-Trees, the dataset is bisected at the median point along the dimension of highest variance, forming two geometric axis-aligned child regions, which are subsequently split using the same logic, and so on, to form a deterministic tree.  For search, deterministic trees may scale poorly with dimensionality.  Several randomization techniques may be applied to the former strategy, which results in a non-deterministic tree.  We use a standard implementation of FLANN \cite{muja2009fast}, in which the split point at each level is chosen randomly among those dimensions that exhibit the greatest variance.  A constant number of trees (we use 10 trees in experiments) are built using independent random choices of the split dimension, and all trees are searched for each query.

\section{Method}
\label{method}
\begin{figure}
\centering
\includegraphics[scale=0.5]{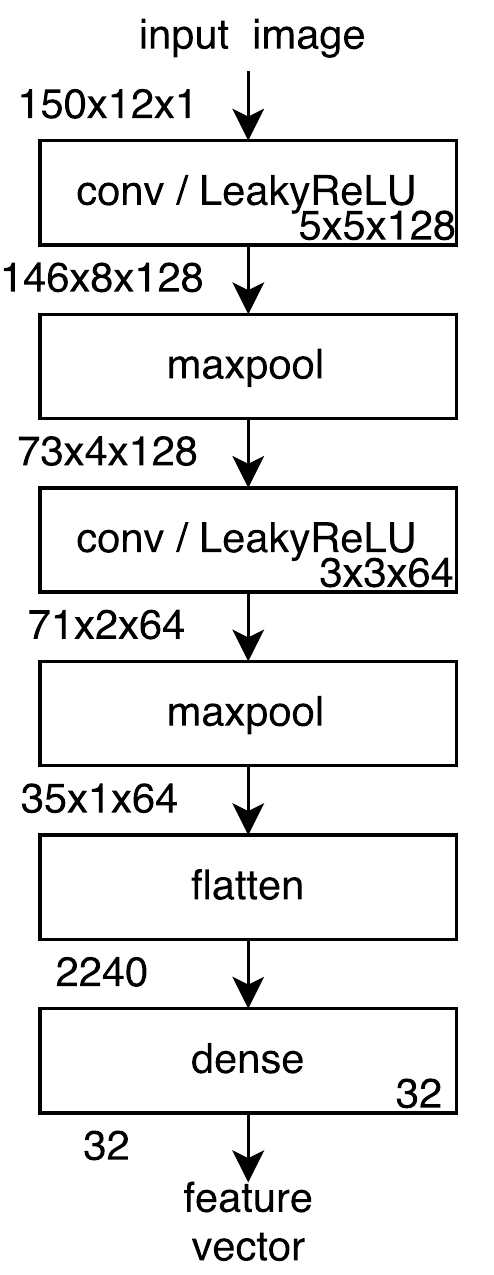}
\caption{Neural network trained to produce similarity features from image-rendered string queries}
\label{fig:network}
\end{figure}

% describe big picture process for detecting spoofed names
We utilize a Siamese network as a key component for predicting the visual similarity between a query string and a whitelist of potential strings that an attacker may spoof.  Our process includes the following steps for determining whether a query is a possible domain or process name spoof.
\begin{enumerate}
\item A query string is rendered as a binary image to capture its visual representation. Independent of the query, whitelist strings are rendered into images of fixed size using a common font.  
\item From the rendered image, image features are extracted using a neural network, shown in Fig. \ref{fig:network}.  This network was trained in a Siamese architecture to capture visual similarity between image-rendered strings and possible spoofs.  The resulting features are those learned by the Siamese network to best capture image similarity between rendered strings and synthesized spoofs.
\item We query a randomized KD-Tree index for feature vectors with Euclidean distance below a specified threshold to the query feature, and report strings corresponding that correspond to spoofs.
\end{enumerate}

In what follows, we provide additional details about components of this process.

\subsection{Neural network similarity model}
\begin{figure}
\centering
\includegraphics[scale=0.45]{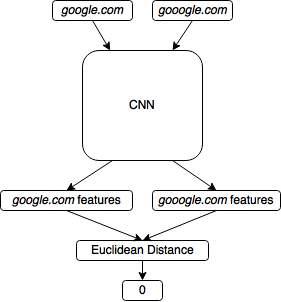}
\caption{Overview example of training the Siamese Neural network.  \textit{google.com} and \textit{gooogle.com} are spoofing pairs and the CNN is trained such that the euclidean distance between their respective features is $0.0$.}
\label{fig:siamese_cnn}
\end{figure}

The neural network in Fig. \ref{fig:network} is intended to produce a feature vector from an input image of rendered text.  In our model, we render images of size 150x12 with white text on black background using \texttt{Arial} TrueType font.  In our experiments, the image size accommodates horizontal space for 25 characters---an artificial limitation that is trivially extended without other dependent changes in the process.

With well-structured input, our network can be relatively small.  We choose two convolution layers with leaky ReLU activations \cite{maas2013rectifier}, each followed by maxpooling with downsampling.  The convolutional layers are followed by a single dense layer that maps the flattened output of the convolutional layers to a 32-dimensional feature vector.

Training the network is using a Siamese architecture in the normal way: a pair of input images $\left(\mathbf{x}_1, \mathbf{x}_2 \right)$ is compared via Euclidean distance in (\ref{eqn:siamese}) as
$d_{\mathbf{W}}\left( \mathbf{x}_1, \mathbf{x}_2 \right)$, and are penalized via contrastive loss in (\ref{eqn:contrastive}).  Parameters of the network are updated via backpropagation.  In our experiments, we use the RMSProp optimizer on batches of 8 images. An example of the entire Siamese CNN is given in Figure \ref{fig:siamese_cnn}.

\subsection{KD-Tree Index}
\begin{figure}
\centering
\includegraphics[scale=0.5]{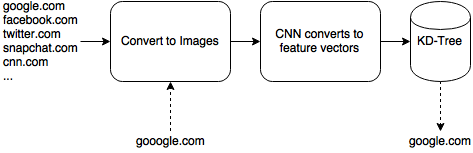}
\caption{Construction of the index takes in many domain (or process) names that are converted to images that are converted to feature vectors by the CNN.  These feature vectors are indexed using KD-Trees.  A potentially malicious name is converted to an image and feature vectors in the same way and used as a query into the KD-Tree index.  Close matches are flagged as spoofing attacks.}
\label{fig:index_construction}
\end{figure}

Potential targets of spoofing attacks are converted to features vectors with the CNN described above.  These feature vectors are indexed using ten randomized KD-Trees, where each tree is grown to purity (1 sample per leaf node).  We perform 128 checks on each query unless otherwise specified.  The KD-Tree implementation in \cite{muja2009flann} is used for experiments in this paper.  Figure \ref{fig:index_construction} demonstrates how a KD-Tree index is constructed and queried.
 
\section{Results}
\label{results}
All experiments are run on two datasets.  The first dataset is constructed using the National Software Reference Library (NSRL) \cite{allen2016national} using all files with the \textit{.exe} and \textit{.dll} and a filename of at least four characters (not including the extension.  Benign pairs (i.e., not spoofing attacks) are created by calculating an all-to-all edit distance and retaining all pairs such that:
\begin{equation}
d(x_1, x_2) \le 3,
\end{equation} 
\noindent where $d$ is the edit distance function (Levenshtein distance).  The edit distance of three is fairly small and chosen to make the dataset one that distinguishes visual similarity from edit distance similarity.  This data sets helps highlight the shortcomings of various algorithms. Malicious pairs (i.e., spoofing attacks) are created by generating spoofing attacks using the file names extracted from NSRL.  Spoofing attacks are generated using thousands of character swaps using both ASCII and unicode characters.  The second data set is composed similar to the NSRL data set except that it was generated using 100K active web domains.  The restriction on edit distance ($d \le 3$) was removed when generating the domain data set. This was due to a lack of non-spoofing pairs with distance less than four.

Note that benign strings in both data sets are predominantly composed of ASCII characters.  However, this would not be the case when deploying the system in many non-English speaking countries.  For this reason, any work using this dataset should not use the presence of unicode as an indicator of spoofing attacks.

\subsection{Setup}
For both data sets, we randomly partition the data into training, testing and validation sets.  A separate neural network was trained for each data set.  The validation set is used during training to prevent over-fitting.  Efficacy results are calculated using \textit{Area Under the Curve (AUC)} of the \textit{Receiver Operating Characteristic (ROC)}.  

For comparison, the Siamese neural network is compared to two string matching techniques: conventional edit distance and visual edit distance \cite{linari2009typo, EditDistanceVC}.

\subsection{Distance Measure Effectiveness}
\begin{figure}
\centering
\includegraphics[scale=0.35]{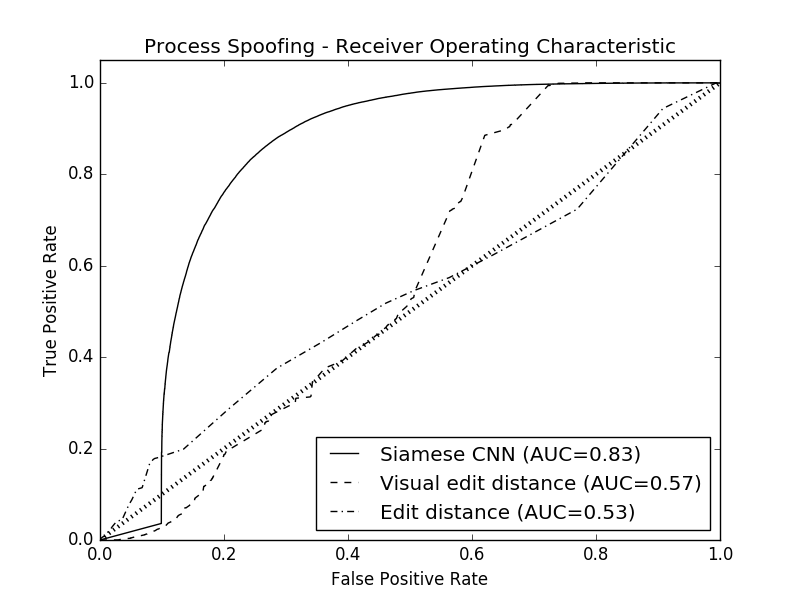}
\caption{ROC curves for classifying process name spoof attacks}
\label{fig:process_roc_curve}
\end{figure}

The first set of experiments compare the effectiveness of the proposed technique to that of conventional edit distance and visual edit distance \cite{linari2009typo, EditDistanceVC}, which we re-implement from descriptions for comparison.  Figure \ref{fig:process_roc_curve} shows the ROC for the process name data set.  Standard edit distance has an ROC very close to $0.5$ (chance).  This is expected as all non-spoofing and spoofing pairs had an edit distance not exceeding $3$, making it difficult to do significantly better than chance using edit distance alone.  Surprisingly, visual edit distance is only slightly improved over edit distance. Lack of improvements highlights the difficulty of manually curating distance measures.  The number of possible characters is extremely large when including unicode and manually deriving all to all distances from each character is unfeasible.  One could attempt to learn an all to all distance between characters, but manually creating such a data set to learn on is also prohibitively expensive.

\begin{figure}
\centering
\includegraphics[scale=0.35]{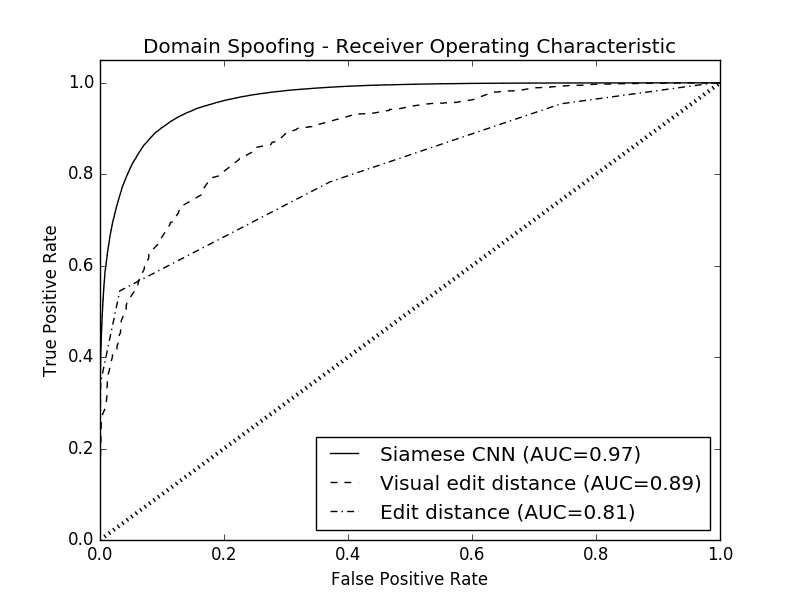}
\caption{ROC curves for classifying domain name spoof attacks}
\label{fig:domain_roc_curve}
\end{figure}

Figure \ref{fig:domain_roc_curve} shows the ROC curve for the domain data set.  Note that all three methods perform far better than the process name data set due to non-spoofing pairs having edit distances that are greater than $3$.  However, the CNN performs significantly better than the other two techniques.  As expected, the visual edit distance is improved over the standard edit distance.

\subsection{KDTree Performance}
The second set of experiments measures the speed improvements and recall degradation when using a KDTrees to index features derived from our model.  The KDTree is used to index known strings that may be spoofed.  For example, the top 100K most visited domains can be converted to feature vectors using the model and indexed as possible targets for homoglyph attacks.  When a new domain is seen, it is converted to a feature vector using the model and is compared to everything in our index.  A naive linear scan will take $nd$ computations where $n$ is the number of elements in our index and $d$ is the number of dimensions.  On the other hand, a KDTree index will only take $c \times \left(\log\left(n\right) + d\right)$ where $c$ is the number of checks used by the KDTree.  (The number of checks is the number of leaf nodes visited in the search.)  We use $c=128$, and in practice $c$ is typically on the order of 64 to 256 making a KDTree far faster than a naive linear scan for large data sizes.  However, this speed increase comes at a cost of lower recall. 

\begin{figure}
\centering
\includegraphics[scale=0.30]{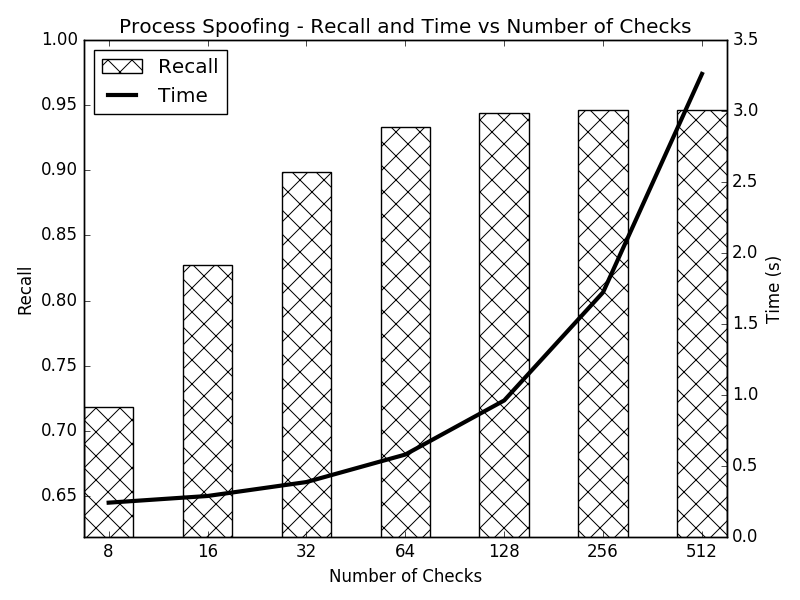}
\caption{Displays the tradeoff of speed and recall with varying number of checks.}
\label{fig:process_recall_curve}
\end{figure}

\begin{figure}
\centering
\includegraphics[scale=0.30]{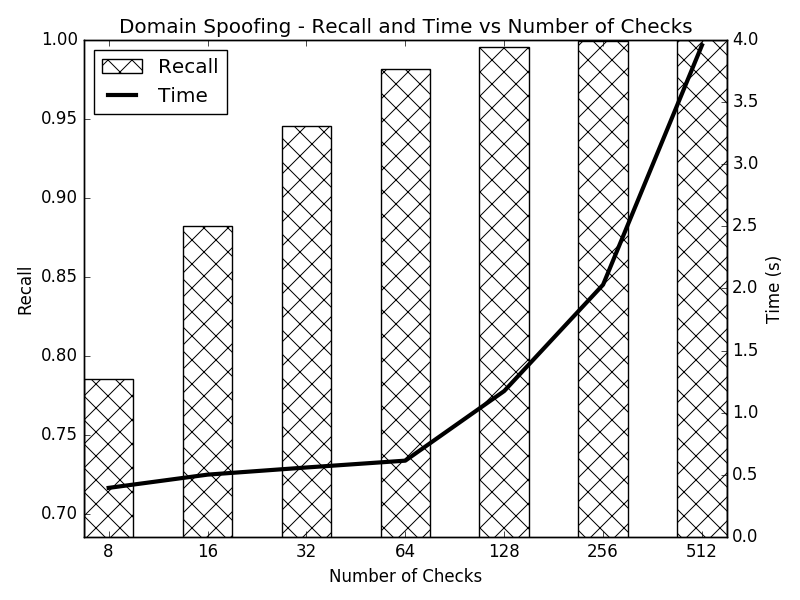}
\caption{Displays the tradeoff of speed and recall with varying number of checks.}
\label{fig:domain_recall_curve}
\end{figure}

Figure \ref{fig:process_recall_curve} and Figure \ref{fig:domain_recall_curve} displays the tradeoff between speed and recall with increasing number of checks for the process data and domain data respectively.  This experiment is run on 50,000 indexed elements and 50,000 queries.  The number of checks equates to the number of leaves explored in a search for the nearest neighbor.  The closest element in each explored leaf is returned as the nearest neighbor.  The likelihood of finding the true nearest neighbor increases with the number of leaves explored. However, the time it takes to search also increases.

\begin{figure}
\centering
\includegraphics[scale=0.3]{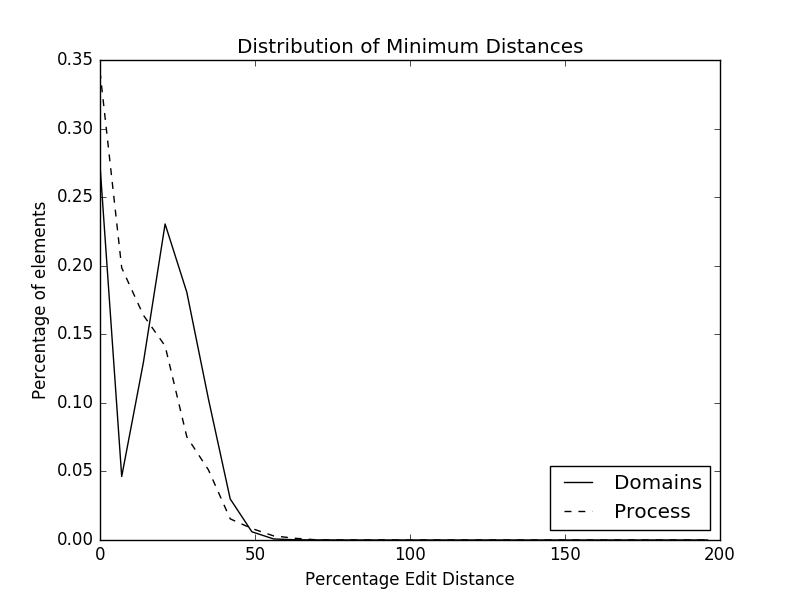}
\caption{Displays the distribution of distances from each process/domain name to its nearest neighbor.  Distance is defined as the percentage edit distance (i.e., the edit distance normalized by the string length).}
\label{fig:percent_edit_distance}
\end{figure}

There is one main differences between the performance of the two data sets.  The domain data set achieves near 1.0 recall while the process data set achieves near 0.95 recall.  One cause of degradation in the process name data set are clusters of very similar process names.  For example, some files in the NSRL data set have versioning information in their file names (e.g., \textit{firefox-1.5.0.1.tar} and \textit{firefox-2.0.0.1.tar}).  Each element in these clusters will have very similar feature vectors generated by the CNN making it more likely for a KD-Tree to return incorrect results.  Figure \ref{fig:percent_edit_distance} shows the distribution of distances from each process/domain name to its nearest neighbor.  Distances are calculated using the edit distance normalized by the string length.  

As can be seen in Figure \ref{fig:percent_edit_distance}, process names have a much larger percentage of nearest neighbors falling in the sub 10\% range than the domain dataset.  This distribution of data can degrade performance as seen in Figure \ref{fig:process_recall_curve}.

Both datasets produce nearly identical runtime behavior, and get the best recall/time trade-off with 128 checks.  The number of checks was based on 50,000 elements and is expected to increase with the number of elements in the index.

\subsection{Visualizing nearest neighbors}
\begin{figure}
\centering
\includegraphics[scale=0.4]{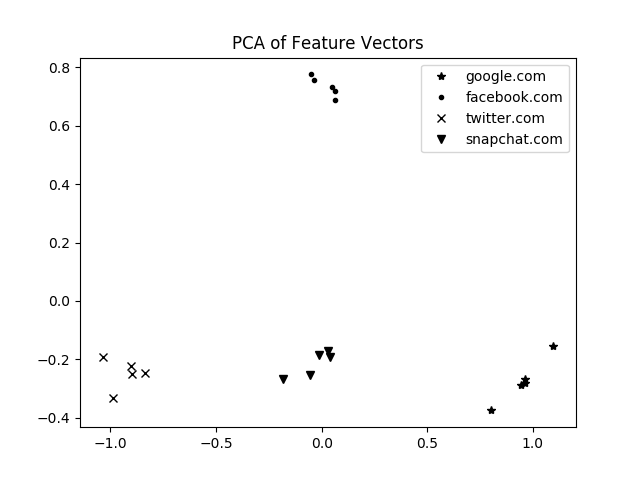}
\includegraphics[scale=0.6]{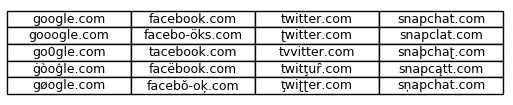}
\caption{(top) Two dimensional PCA projection of feature vectors derived from \textit{google.com}, \textit{facebook.com}, \textit{twitter.com}, and \textit{snapchat.com} along with 4 homoglyph attacks; (bottom) homoglyphs for each of the domain names.}
\label{fig:pca}
\end{figure}

Figure \ref{fig:pca} displays the feature vectors of twenty domain names, consisting of 4 domain names each with 4 additional homoglyphs.  A PCA projection is performed on these feature vectors to reduce the number of dimensions to two. The names consist of \textit{google.com}, \textit{facebook.com}, \textit{twitter.com}, and \textit{snapchat.com} along with four homoglyph attacks for each domain.  Note how each domain and respective homoglyph attacks cluster tightly demonstrating that our learned feature vectors are able to distinguish well between domain names.  Distinguishability allows us to predict  spoofing attacks with very low false positive rates.

\section{Conclusion}
\label{conclusion}
We presented a technique\footnote{Code and data are publicly available at \REPO.} for detecting domain and process homoglyph attacks using a Siamese CNN.  Names are converted to images and passed to the CNN to convert the name to a feature vector.  The CNN is trained such that similar strings (i.e., spoofing attacks) generate feature vectors that have a small Euclidean distance while dissimilar strings produce feature vectors that have a large Euclidean distance.  Results were compared to conventional detection methods using edit distance and demonstrated a 13\% to 45\% improvement in terms of area under the ROC curve. 

%This paper also assessed the ability of feature vectors to be indexed using a KD-Tree and showed that proper approximate indexing can result in fast and accurate spoofing detection with little degradation from exact matches.  This characteristic allows this technique to be run in real world systems. Finally, all code and and data sets were made publicly available such that experiments can be reproduced.  We encourage the use of data sets to foster research in the area.

%\newpage
\IEEEtriggeratref{14}
\bibliographystyle{IEEEtranS}
\bibliography{aisec}



%\bibliographystyle{IEEEtran}

\end{document}